\documentclass[11pt]{article}
\usepackage{color,amsfonts,amsmath,amssymb,mathrsfs,verbatim,epsf}

\def\inn{\in}

\def\TM{\mbox{\it TM}}

\def\rrr{{\mathbb{R}}}
\def\ccc{{\mathbb{C}}}

\def\ooo{{\mathbb{O}}}

\def\b1{\mbox{\boldmath $1$}}

\def\v{\mbox{\boldmath $v$}}

\def\bh{\mbox{\boldmath $h$}}

\def\bv{\mbox{\boldmath $v$}}

\def\hG{\hat{G}}

\def\lvtfep{L_8(\v_{248})_{\mathrm{E}_8}=1}

\def\lpvng{L_p(\v_n)_{\hG}=1}

\def\htwc{\mbox{h}_2\ccc}
\def\hthc{\mbox{h}_3\ccc}

\def\htwo{\mbox{h}_2\ooo}
\def\htho{\mbox{h}_3\ooo}

\def\sltc{\mbox{SL}(2,\ccc)}

\def\slthc{\mbox{SL}(3,\ccc)}

\def\sltho{\mbox{SL}(3,\ooo)}

\def\uo{\mbox{U}(1)}

\def\sutw{\mbox{SU}(2)}

\def\suth{\mbox{SU}(3)}

\def\soot{\mbox{SO}^+(1,3)}

\def\sootnm{\mbox{SO}^+(1,n-1)}

\def\gt{\mbox{G}_2}
\def\ff{\mbox{F}_4}
\def\ee{\mbox{E}_8}

\def\ese{\mbox{E}_7}

\def\esi{\mbox{E}_6}

\def\mcX{{\mathcal X}}
\def\mcY{{\mathcal Y}}

\def\mmin{m_{\mathrm{min}}}
\def\mtot{m_{\mathrm{tot}}}

\def\gpath{ }

\def\setb{\setlength{\baselineskip}{0.625\baselineskip}}

\begin{document} 

{\setlength{\baselineskip}{0.625\baselineskip}

\begin{center}
 
{\LARGE{\bf  From Unified Field Theory to the}} \\
 \vspace{4pt}
 {\LARGE{\bf  Standard Model and Beyond}} \\ 
 
   \vspace{6pt}

 \mbox {{\Large David J. Jackson}}\footnote{email: david.jackson.th@gmail.com}  \\

  \vspace{4pt}
 
 { \large September 12, 2018}

 \vspace{3pt}

{\bf  Abstract}

\vspace{-10pt} 
 
\end{center}

    One hundred years ago this year attempts began to generalise general relativity with the ambition of 
incorporating electromagnetism alongside gravitation in a unified field theory. These developments led to 
gauge theories and models with extra spatial dimensions that have greatly influenced the modern-day pursuit 
of a unification scheme incorporating the Standard Model of particle physics, again ideally together with 
gravity. In this paper we motivate a further natural generalisation from  extra spatial dimensions at an 
elementary level which is found to much more directly accommodate distinctive features of the Standard 
Model. We also investigate the potential to uncover new physical phenomena, making a case in the neutrino 
sector for one left-handed neutrino state to be massless, and emphasise the opportunity for a close 
collaboration between theory and experiment. The new theory possesses
  a very simple interpretation regarding the underlying source of these empirical structures. 
{

\vspace{-16pt}
\tableofcontents
\vspace{-11pt}

}



\section{Introduction: Unified Field Theories}
\label{una1}

  The now familiar pattern of elementary particle multiplets consisting of leptons, 
quarks, gauge bosons and the Higgs was termed the `Standard Model' in the mid-1970s 
(\cite{Pais2} chapter~21(e)).
  The richness of this structure contrasts with the situation
 a hundred years ago, during the era of the `Bohr atom', when the basic constituents 
of particle physics were simply  the electron, the proton (identified with the 
hydrogen nucleus) and the photon; with electromagnetism and gravitation being the only 
two known fundamental forces (\cite{Pais} chapter~17(a)).
 Early unified field theories were based on \mbox{generalisations} of Einstein's 
theory of gravity, then only recently expounded~\cite{Eingr}, which itself possesses a 
simple geometric interpretation.

  For general relativity the 
 \textit{assumption} of an extended globally flat 4-dimensional Minkowski spacetime,  
the original arena of special relativity, was \textit{dropped}.
  The theory was motivated by the apparent absence of the force of gravity in a local 
inertial reference frame, with special relativity still holding for all 
non-gravitational physics within such a local `free fall' arena by the equivalence 
principle (\cite{Pais} chapter~9). 
  An infinitesimal `proper time' interval $\delta s$ associated with any free fall 
trajectory at any location can by definition be expressed in a local inertial 
reference frame as:
\begin{equation}
   (\delta s)^2 =  (\delta x^0)^2 - (\delta x^1)^2 - (\delta x^2)^2
    - (\delta x^3)^2  \; = \; \eta_{ab}\delta x^a \delta x^b \label{sfourd} 
\end{equation}  
  in terms of local coordinates $\{x^a\}$, with the Lorentz metric
  $\eta = \mbox{diag}(+1,-1,-1,-1)$ and $a,b = 0,1,2,3$, in a form invariant under 
local Lorentz  transformations between inertial frames. 
 With respect to extended general coordinates $\{x^{\mu}\}$, with $\mu,\nu= 0,1,2,3$, 
a local proper time interval can be expressed as an invariant under general coordinate 
transformations with:
\begin{equation}
   (\delta s)^2  \; = \; g_{\mu\nu}(x)\delta x^{\mu} \delta x^{\nu} \label{gfourd} 
\end{equation}  

     The force of gravity is  ascribed to the metric field function $g_{\mu\nu}(x)$ 
which generalises $\eta_{ab}$ on the global scale and describes the geometry of an 
extended curved spacetime.
  It was then natural to enquire whether a further generalisation from general 
relativity might also provide an \textit{explanation} of electromagnetic phenomena on 
relaxing further assumptions regarding the 4-dimensional spacetime metric geometry.  
  In 1918 the first attempt was made by Weyl on considering the path-independence of 
the length of a parallel transported 4-vector
 to be a last vestige of rigid Euclidean geometry in general relativity.
On dropping that assumption and introducing a scale factor Weyl developed a theory  
 of electromagnetism of geometric origin~(\cite{Weyl1}, \cite{ORaif} chapters 1--3).
 
  While that theory was flawed, by 1929 the scale factor applied to the metric of 
general relativity was converted into a phase factor applied to a complex wavefunction 
in quantum mechanics, successfully describing a theory of 
electromagnetism~(\cite{Weyl3}, \cite{ORaif} chapters~4--5). This `gauge theory' for 
electromagnetism with gauge group $\uo$ was generalised for larger non-Abelian gauge 
symmetries in the 1950s~(\cite{ORaif} chapters~8--10) and became central to the 
modern-day structure of the Standard Model and Grand Unified Theories.
 A degree of success has been achieved for unification groups 
  such as $\esi$, $\ese$ and $\ee$ (see for example~\cite{Gur1,Gur7}). However, 
 while these exceptional Lie groups are also of particular interest owing to the high 
degree of symmetry they describe and the uniqueness of these mathematical structures, 
a \textit{clear underlying conceptual motivation}, whether geometric or otherwise, for 
their application in particle physics is still lacking.

In the 1920s the unified field theory of Kaluza~\cite{Kaluza} and Klein~\cite{Klein} 
was also introduced  (\cite{Pais} chapter~17(c), \cite{ORaif} chapter~3),
 with the assumption that spacetime should be limited to the 4-dimensional arena of 
general relativity being dropped. While various equations of electrodynamics could be 
extracted from a 5-dimensional spacetime framework, providing an element of formal 
geometric unification with general relativity,
 no new phenomena were predicted. 
   Nevertheless the elegance and unity of the Kaluza-Klein idea,
  together with the realisation that the geometry of extended spacetimes with 
 further extra spatial dimensions 
   could be adapted to incorporate non-Abelian gauge theory (see for 
example~\cite{Cho}),
  has motivated subsequent unification schemes. 
Since the 1970s the ambition has been to accommodate structures of the Standard Model 
of particle physics, or even a Grand Unified Theory, via the properties of the extra 
spatial dimensions over 4-dimensional spacetime.

\section{Elementary Extra Spatial Dimensions}
\label{una2}

 At the most elementary level of purely local structure adding extra \textit{spatial} 
dimensions implies augmenting the quadratic expression for the 4-dimensional proper 
time interval $\delta s$ of equation~\ref{sfourd} to the $n$-dimensional form:
\begin{equation}
   (\delta s)^2  = (\delta x^0)^2 - (\delta x^1)^2 - (\delta x^2)^2 - (\delta x^3)^2
   \, - \, (\delta x^4)^2 \ldots\ldots - (\delta x^{n-1})^2   = 
             \hat{\eta}_{ab}\delta x^a \delta x^b  \;
  \label{snd}
\end{equation} 
  where $\{x^4, \ldots, x^{n-1}\}$ are the $(n-4)$ extra dimensions, 
  $\hat{\eta} = \mbox{diag}(+1,-1, \ldots, -1)$ is the extended local Lorentz metric 
and $a,b = 0,\ldots,(n-1)$. On dividing both sides by $(\delta s)^2$ 
   and defining the components $v^a = \frac{\delta x^a}{\delta s}$ in the limit 
    $\delta s \to 0$
   this expression can be written as:
\begin{equation}
  \vert \bv_n \vert^2  := \;\! (v^0)^2 - (v^1)^2 - (v^2)^2 - (v^3)^2
   \, - \, (v^4)^2 \ldots\ldots - (v^{n-1})^2 \;\!  = \, 
             \hat{\eta}_{ab}v^a v^b  \, = \, 1
  \label{vnd}
\end{equation} 
  in terms of the components  of the `$n$-velocity' vector
    $\bv_n = (v^0, \ldots, v^{n-1}) \in \rrr^n$.
  The simplest and most direct means of constructing a physical theory based on this 
structure is to assume the breaking of the $\sootnm$ symmetry of equation~\ref{vnd} in 
projecting the first four components 
 onto the local tangent space, $\bv_4 = (v^0,v^1,v^2,v^3) \in \TM_4$, at any location 
on the 4-dimensional spacetime manifold $M_4$, upon which a preferred local 
\textit{external} Lorentz $\soot \subset \sootnm$ symmetry acts. On taking the 
residual components of equation~\ref{vnd} to form the basis for `matter fields' 
\textit{in} the extended spacetime we directly deduce the following symmetry breaking 
pattern:
\begin{eqnarray}
        \sootnm & \to &  \soot \; \times \;  \mbox{SO}(n-4)    \label{sosb} \\
	    \bv_4 \in \rrr^4  \:\, & : &  \,  \mbox{4-vector} 
		 \qquad\quad \!\!\! 
		     \mbox{invariant}   \qquad \mbox{: tangent vector}   \nonumber  \\
 \raisebox{0pt}[0pt][0pt]{ {\raisebox{+1.7ex}{$\bv_n \to 
  \left\{ \begin{array}{c}  \\ \vspace{-12pt}  \end{array}  \right. 
   \!\!\!\!\!\! $}} }		 
	 \underline{\bv}_{n-4} \in \rrr^{n-4} \!\!\!\!  & : & \;\:\,  \mbox{scalar} 
	    \qquad\;\!\!\!  
		   (n-4)\mbox{-vector} \quad\hspace{-0.3pt}  \mbox{: matter field}   
	     \label{vnbits}
\end{eqnarray}   

   In this simple picture the matter field $\underline{\bv}_{n-4}(x)$  in spacetime 
$M_4$, as a Lorentz scalar that transforms under the $(n-4)$-dimensional vector 
 representation of the residual \textit{internal} gauge symmetry SO$(n-4)$, does not 
remotely resemble structures of the Standard Model of particle physics for any value 
of $n$.
Rather than specifically \textit{adding} more sophisticated structures we shall 
motivate an \textit{intrinsic} generalisation that greatly improves this situation.

\section{Extra Dimensions Reconsidered}
\label{una3}

  Here we observe that
 since we do not perceive or navigate around the extra dimensions there is no 
compelling reason for the additional components to possess the local structure of 
 $\{x^4, \ldots, x^{n-1} \}$  in equation~\ref{snd} as a \textit{quadratic} extension 
to the local 4-dimensional spacetime form of equation~\ref{sfourd} (with the minus 
signs from the Lorentz metric signature convention).
  That is, the extra components in equation~\ref{snd}  have the `spatial' property of     
 adding quadratically to form local `lengths' $\delta \Sigma$, with for example:
\begin{equation} 
 \label{pythag}
   (\delta\Sigma)^2 \, = \, (\delta x^4)^2 \, + \, (\delta x^{n-1})^2
\end{equation}   
    which via the Pythagorean theorem describes right-angled triangle structures as a 
basis for a local Euclidean geometry. This property is only \textit{required} for the 
local geometric structure of 
  $\{\delta x^1, \delta x^2, \delta x^3\}$  in forming the basis of the extended 
\textit{external} 3-dimensional space that we do perceive and move around in,
 while the  \textit{assumption}
  of this locally Euclidean form can be \textit{dropped} for
  the extra components.

  This unnecessary restriction seems even more artificial on considering
  \mbox{large $n$} since then \textit{almost all} of the components on the right-hand 
side of equation~\ref{snd} are not required to take a quadratic form as the $\{\delta 
x^a\}$ for all $a>3$ do not represent a physical perceived space. 
  However the left-hand side of equation~\ref{snd} still describes a simple interval 
of proper time $(\delta s)$, invariant under 
 SO$^+(1,n-1)$ transformations, 
  which is hence pivotal in threading together all of the
  basis-dependent components on the right-hand side and in defining this structure. In 
fact we can interpret equation~\ref{snd} as simply representing a possible arithmetic 
expression 
\textit{for} a real proper time interval $\delta s \in \rrr$ and
 then ask what further possibilities there may be.

 While intervals of time add linearly, 
   as objectively recordable by a clock,
  on exploiting the basic arithmetic properties of the real numbers expressions
   for an infinitesimal interval
   can be written down for $(\delta s)$,
  $(\delta s)^2$, $(\delta s)^3$, \ldots or $(\delta s)^p$ in general, for 
$p=1,2,3,\ldots$, of which equation~\ref{snd} represents only a particular case for 
$p=2$.
 This suggests that the functional form on the right-hand side of equation~\ref{snd} 
can be generalised to a $p^{\mathrm{th}}$-order homogeneous polynomial expression 
  in $n$ components $\{\delta x^a\}$, with $a,b,c = 0, \ldots, n-1$:
\begin{equation}
 \label{salpha}
  (\delta s)^p  \; = \; \alpha_{abc\ldots}\delta x^a \delta x^b \delta x^c \ldots
    \quad \mbox{with each} \quad \alpha_{abc\ldots} \inn \{-1,0,1\}
\end{equation}
   \textit{provided} we can extract an appropriate 4-dimensional quadratic 
substructure
  in four components $\{\delta x^0, \delta x^1,\delta x^2, \delta x^3 \}$, in the form 
of the right-hand side of equation~\ref{sfourd}, as required
       to represent the local geometric structure of the external spacetime $M_4$. 
	   That is, we require that equation~\ref{salpha} can in general be written in the 
form:
 \begin{equation}
 \label{sfourp}
  (\delta s)^p  \; = \; \left[
    \eta_{ab}\delta x^a \delta x^b \right]  (\delta x^4, \ldots ,\delta x^{n-1})^{p-2}
   \; + \;  (\delta x^0, \ldots ,\delta x^{n-1})^{p} 
\end{equation}
  where here in the first term $a,b = 0,1,2,3$ in the first factor and the second 
factor represents a $(p-2)^{\mathrm{th}}$-order polynomial in the remaining $(n-4)$ 
components, while the second term represents the further $p^{\mathrm{th}}$-order 
polynomial contributions to equation~\ref{salpha}.

In order to establish a convenient notation and avoid expressions with infinitesimal 
elements we can in turn generalise equation~\ref{vnd} by again defining an $n$-vector 
$\bv_n \in \rrr^n$ with the generally finite components
  $v^a = \frac{\delta x^a}{\delta s} 
          {\big{\vert}}_{\mbox {\tiny $\delta s \! \to \! 0$}}$, 
 and on dividing both sides of equation~\ref{salpha} by $(\delta s)^p$ we define:
\begin{equation}
  \label{lpvn}
  L_p(\bv_n)_{\hat{G}} \; := \; 
 \alpha_{abc\ldots} \frac{\delta x^a \delta x^b \delta x^c \ldots} 
                           {\delta s \;\;\delta s \;\;\delta s \,\ldots} 
			\Big\vert_{\delta s \to 0}\
  \;  = \; 	
   \alpha_{abc\ldots}v^a v^b v^c \ldots \; = \; 1
\end{equation}
 again with $a,b,c = 0, \ldots, n-1$ and each $\alpha_{abc\ldots} \inn \{-1,0,1\}$. 
Here $L_p$ denotes a \mbox{$p^{\mathrm{th}}$-order} homogeneous polynomial function of 
the $n$ components of $\bv_n$ with the full symmetry \mbox{group $\hat{G}$}.
 For $p>2$ particular values for $p$ and $n$ will be \textit{inherently} preferred as 
unique mathematical structures which  possess a high degree of symmetry, while 
subsuming equation~\ref{sfourd}, will be highlighted. In this sense the progression 
from `\mbox{spacetime} forms' to `forms of time' is both more general and yet more 
restrained, and in a manner that leads to well known unification groups as we describe 
below.


\section{Utilising the Form of Time}
\label{una4}

As a means of explicitly embedding the 4-dimensional quadratic spacetime form inside a 
higher-order homogeneous polynomial form of time we rearrange equation~\ref{sfourd} in 
the fashion of equation~\ref{lpvn} and
 write the resulting expression as the determinant of a $2 \times 2$ Hermitian complex 
matrix:
\begin{equation}
  \label{lqdet}
  L_2(\bv_4)_{\mathrm{SL}(2,\ccc)} \; = \; \eta_{ab}v^a v^b
 \; = \;    \det (\bh)  \; = \;    \det \left( \!\!
	   \begin{array}{cc} v^0 + v^3 & v^1 - v^2i \\
		   v^1 + v^2i  & v^0 - v^3  \end{array}  \!\! \right)
		   \; = \; 1
\end{equation}  
  with the Lorentz 4-vector $\bv_4 \equiv \bh \in \htwc$. As indicated 
  this determinant form is  invariant under the action of the symmetry group $\sltc$ 
as the double cover of the Lorentz group $\soot$ (\cite{KKone} equations~16 and 17).
  This 4-dimensional form can be embedded directly within the determinant of a $3 
\times 3$ Hermitian complex matrix, which we interpret as a \textit{cubic} form of 
time in \textit{nine} components consistent with equation~\ref{lpvn}, now with an 
augmented $\slthc$ symmetry:
\begin{eqnarray}
   L_3(\bv_9)_{\mathrm{SL}(3,\ccc)} \!\!\! & = & \!\!\!  \det  \!
	\left( \! \begin{array}{cc|c}  
	  v^0  +  v^3  
   &  v^1  -  v^2 i  
   &  v^4  +  v^5 i    \\
	  v^1  +  v^2 i 
   &  v^0  -  v^3 
   &  v^6  +  v^7 i   \\  \hline
      v^4  -  v^5 i 
   &  v^6  -  v^7 i  
   &  v^8     \end{array} \! \right)  =   \det \!
   \left( \! \begin{array}{c|c} 
        \,\,\, \bh \!\!     \begin{array}{cc} &  \\  &  \end{array} \!\!\!   &
        \:\!  \psi \!\!\!\;  \begin{array}{cc} &  \\  &  \end{array}
		    \!\!\!\!\!\!\!\!\!\! 
				                      \\ \hline
        \,\,\,\,\;\! \psi^{\dagger} \!\!\!\! \begin{array}{cc}  
		        &   \end{array}   &	  
		   n \!      \end{array} \!  \right)  = 1  \qquad  \label{lvni} 
\end{eqnarray}	 
\begin{equation} 
  = \;  \det(\bv_9) \;  = \;
                 \Big[ \eta_{ab}v^a v^b   \Big] n 			   
		\,  - \,  2\bh \!\cdot\! (\psi\psi^{\dagger})  \, = \,  1 
		 \qquad \qquad \quad \;\;\;\!   \label{lqinc}
\end{equation}
   with $\bv_9 \in \hthc$, $\bh \inn \htwc$, $\psi \inn \ccc^2$ and here $n=v^8 \in 
\rrr$ while $a,b = 0,1,2,3$. In the final term $\bh \!\cdot\! (\psi\psi^{\dagger})$ is 
the Lorentz inner product between the 4-vectors associated with 
  $\bh, \psi\psi^{\dagger} \in \htwc$ (\cite{TimeE} equations~23 and 70). 

  The full $\slthc$ symmetry of $L_3(\bv_9)_{\mathrm{SL}(3,\ccc)} = 1$ is broken 
   through a preferred \textit{external} 
$\sltc \subset \slthc$ symmetry acting upon a necessary choice of
   $\bv_4 = (v^0,v^1,v^2,v^3) \in \TM_4$ subcomponents of
 $\bv_9 \in \hthc$ projected onto a local inertial reference frame from 
equations~\ref{lvni}--\ref{lqinc}.  
The extraction of this necessarily \textit{quadratic} substructure 
  to match the local geometry of the external spacetime,
   via the square brackets in equation~\ref{lqinc},
 also leaves a 
 residual  \textit{internal} $\uo \subset \slthc$ symmetry 
 that can be interpreted as a gauge group underlying a theory of electromagnetism    
  (\cite{KKone} subsections~2.3 and 4.2).
The broken symmetry reduces the full 9-dimensional vector $\bv_9 \in \hthc$ to the 
three parts introduced in equation~\ref{lvni} with 
 the Lorentz $\sltc$ and internal $\uo$ factors  acting upon these subcomponents as:
\begin{eqnarray}
        \slthc & \to &  \sltc \; \times  \; \uo    \label{slsb} \\
	    \bh \inn \htwc  \!\!\!\!\! & : &  \,  \, \mbox{vector} \!
		  \qquad\;\;\;\!\hspace{0.1pt} \;\;   0 
		    \quad\; \mbox{: tangent vector} \nonumber  \\
  \raisebox{0pt}[0pt][0pt]{ {\raisebox{+0.0ex}{$\bv_9 \to 
  \left\{ \begin{array}{c} \\ \\ \vspace{-15pt}  \end{array}  \right. 
   \!\!\!\! $}} }					
	   \psi\inn \ccc^2 \!\!\!  & : & \!\! \,  \mbox{$L$-spinor} \! \qquad\,  \;\;   1 
	               \quad \; \hspace{-0.1pt} \mbox{: matter field}   \nonumber  \\
	    n\inn \rrr    \!   & : &  \; \, \mbox{scalar}\!  \qquad\;\;\,  \;\;   0   
		           \quad\; \mbox{: matter field}   \label{slbits}
\end{eqnarray}  
with the 2-component Weyl spinor $\psi$ taken to be left-handed by convention. 
 Hence the internal $\uo$ gauge symmetry of electromagnetism  also acts non-trivially 
upon a spin-$\frac{1}{2}$ field $\psi(x)$ in spacetime, as indicated by the normalised 
unit charge `1' in equation~\ref{slbits}, while $n(x)$ is a neutral scalar.

 Being central to the symmetry breaking, and having a scalar magnitude
  $\vert \bh \vert = \sqrt{\det(\bh)}$ in the projection onto $\TM_4$, the components 
of the vector field $\bh(x) \in \htwc$ of equation~\ref{slbits} are associated with a 
non-standard Higgs in this theory (also for the further reasons reviewed in 
\cite{TimeE} after figure~4).
In general
the symmetry breaking projection of $\bh\equiv \bv_4 \in \TM_4$ out of the full set of 
components for the $n$-dimensional form of equation~\ref{lpvn} breaks the  original 
full symmetry $\hG$ down to the subgroup:
\begin{equation}
 \mbox{Lorentz} \times G \subset \hG
  \label{gbreak}
\end{equation} 
  where the Lorentz symmetry may be the double cover $\sltc$
   and  $G$ is the internal symmetry, with $G=\uo$ in equation~\ref{slsb}.
  The components of $\bv_n \inn \rrr^n$ are partitioned into subsets that transform 
under irreducible representations of this subgroup, as listed for example in 
equation~\ref{slbits} (and discussed in \cite{KKone} for equation~23 there).
At the same time the set of terms in the expansion of the corresponding form
  \mbox{$\lpvng$}, which is invariant under $\hG$, will be partitioned into subsets 
invariant under the $\mbox{Lorentz} \times G$ broken symmetry of 
equation~\ref{gbreak}, as for the two parts of equation~\ref{lqinc}. 
 For the full theory
 such individually invariant parts of $\lpvng$ which contain a factor of $\bh$, or a 
scalar composition such as $\vert \bh \vert$,   
   are proposed to be associated with `mass terms' in an effective Lagrangian deriving 
from the theory, in part motivating the
 kernel symbol `$L$' in equation~\ref{lpvn}.

   The identification of the local geometric structure of the spacetime manifold  
itself with a quadratic substructure extracted from equation~\ref{lpvn} implies   the 
complete distinction between the external and internal components.  
 Hence the full symmetry $\hG$ of equation~\ref{lpvn}, with which we begin in the 
mathematics of the theory,  is broken \textit{absolutely} to the product of the 
external Lorentz and internal $G$ symmetry in equation~\ref{gbreak} to describe all 
physics that can be defined in 4-dimensional spacetime, consistent with the demands of 
the Coleman-Mandula theorem ultimately for the relativistic quantum theory limit 
(\cite{KKone} subsection~5.3).  

\section{$\esi$, $\ese$, $\ee$ and the Standard Model}
\label{una5}

   While the Lie algebras, including the five exceptional cases of $\gt$, $\ff$, 
$\esi$, $\ese$ and $\ee$, were classified by Killing and Cartan in the late 
$19^{\mathrm{th}}$ century (\cite{Baez1} section~4 opening) an understanding of 
explicit expressions for certain representations of these algebras developed from the 
mid-$20^{\mathrm{th}}$ and continues into the $21^{\mathrm{st}}$ century. For example 
 the smallest non-trivial representation of $\esi$ can be expressed by the space of
 $3 \times 3$ Hermitian \mbox{octonion} matrices $\htho$, as employed in 
1950~\cite{Chev}, with the corresponding determinant preserving  $\esi \equiv \sltho$ 
group action described in explicit detail more recently~\cite{Wang2}.

  In the context of the present theory, while we obtained equation~\ref{lvni} from 
equation~\ref{lqdet} by a natural minimal extension from the $2 \times 2$ to the 
$3\times 3$ matrix case, we can also augment equation~\ref{lvni} by a natural 
generalisation from the complex numbers $\ccc$ to the octonions $\ooo$, which with 
eight real components uniquely form the largest division algebra~(\cite{Baez1} 
sections~1 and 1.1), to obtain the cubic 27-dimensional form:
\begin{equation}
  \label{lvts}
   L_3(\bv_{27})_{\mathrm{E}_6} = \det(\bv_{27}) \,  = \,   \det \!
   \left( \! \begin{array}{c|c} 
        \,\,\,\,\,\, X \!     \begin{array}{cc} &  \\  &  \end{array} \!\!\!   &
         \:\!  \theta \!\!\!\;  \begin{array}{cc} &  \\  &  \end{array}
		  \!\!\!\!\!\!\!\!\!\! 
				                      \\ \hline
        \,\,\,\,\;\, \theta^{\dagger} \!\!\!\! \begin{array}{cc}  
		        &   \end{array}   &	  
		   n \!      \end{array} \!  \right)  = 1 
\end{equation}
 with $\bv_{27} \inn \htho$, $X\in \htwo$, $\theta \in \ooo^2$ and again $n\in \rrr$.
  The $\slthc$ symmetry of  equation~\ref{lvni} is augmented correspondingly  to 
$\sltho \equiv \esi$. On identifying  an external symmetry $\sltc \subset \esi$ acting 
upon a projected Lorentz 4-vector \mbox{$\bv_4 \in \TM_4$}, now identified with $\bh 
\in \htwc$ extracted 
 from subcomponents of $X \inn \htwo$ in equation~\ref{lvts},
  a symmetry breaking pattern is determined with
   (\cite{TimeE} subsection~4.2):
\begin{eqnarray}
        \esi & \to &   \sltc \; \times \; \suth_c \; 
		   \times \; \uo_Q            \label{esisb} \\
	     X\inn\htwo \!\!\!\!\! & : &  \!\!\!\!\!\!   
    \left\{  \begin{array}{cccl}
       \, \mbox{vector}\,  \quad\;  & \quad\,  \mathbf{1} \quad\, & 
	                       \quad\;\;  0 & \quad\!  \mbox{: $\nu$-lepton$/\bh$}  \\
	   \, \mbox{scalar}\,  \quad\;  & \quad\, \mathbf{3} \quad\, &
	                       \quad\;\;  \frac{2}{3}  & \quad\!  \mbox{: $u$-quark}
	  \end{array}   \right.                  \nonumber  
	                               \\
\raisebox{0pt}[0pt][0pt]{ {\raisebox{+2.5ex}{$\bv_{27} \to 
  \left\{ \begin{array}{c} \\ \\ \\ \\ \vspace{-3.0ex} \\ \\   \end{array}  \right. 
   \! $}} }	
	     \theta\inn\ooo^2 \!\!\!\!\! & : &  \!\!\!\!\!\!   
    \left\{ \!\! \begin{array}{cccl}
        L\mbox{-spinor}  \quad\!\!\;\hspace{-0.2pt}  & \quad\,  \mathbf{1} \quad\, & 
	                       \quad\;\;  1 & \quad\!   \mbox{: $e$-lepton}  \\
	    L\mbox{-spinor}  \quad\!\!\;\hspace{-0.2pt}  & \quad\, \mathbf{3} \quad\, &
	                       \quad\;\;  \frac{1}{3}  & \quad\! \mbox{: $d$-quark}
	  \end{array}   \right.                  \nonumber                     \\
	       n\inn \rrr   \!\!\!\!\! & : &    \quad\!\!
	      \mbox{scalar}  \qquad \quad\, \,\,\hspace{0.2pt}  \mathbf{1} 
		     \quad\: \hspace{-0.2pt} 
	                       \qquad\;\:   0    \begin{array}{c} \vspace{0pt} \\
						    \end{array}    
						 \label{esibits}           
\end{eqnarray} 

  Through this natural augmentation from equations~\ref{slsb}--\ref{slbits} we hence 
find an internal non-Abelian symmetry, which is identified with the colour gauge group 
$\suth_c$, alongside the original Abelian gauge group of electromagnetism, now denoted 
$\uo_Q$. The pattern of $\uo_Q$ relative charge magnitudes determined and listed in 
equation~\ref{esibits} as aligned with the $\suth_c$ singlets
 $\mathbf{1}$ and triplets $\mathbf{3}$ leads to the provisional `matter field' 
interpretation of the subcomponent decomposition of $\bv_{27} \inn \htho$ under the 
broken symmetry as representing a generation of Standard Model leptons and quarks, as 
listed in the final column of equation~\ref{esibits}. 

However, with respect to the external $\sltc$ symmetry, while the $d$-quark 
 and $e$-lepton states transform uniformly as a set of four 2-component left-handed 
Weyl spinors, 
 the `$u$-quark' states transform as Lorentz scalars and the neutral components most 
naturally associated with the neutrino, with respect to the internal symmetry, are 
incorporated into a Lorentz 4-vector. 
 Compounding these discrepant features, this natural slot for the `$\nu$-lepton' in 
equation~\ref{esibits} is already occupied specifically by the \mbox{4-vector} $\bh 
\inn \htwc$ subcomponent of $X \inn \htwo$, which is projected onto the tangent space 
of the external spacetime and associated with the Higgs as noted for 
equation~\ref{slbits}. 

The smallest non-trivial representation of the next largest exceptional Lie group 
$\ese$ can be described by an action on a related 56-dimensional space~\cite{Freud}, 
which we denote $F(\htho)$, as again has been studied in detail more recently (see for 
example~\cite{Rios}). The $\ese$ action preserves a homogeneous quartic form $q$ on 
the space $F(\htho)$ which, while not expressed as a determinant function itself, 
contains the $\esi$ action on the 27-dimensional cubic form of equation~\ref{lvts} and 
hence can be considered as a further 
natural augmentation consistent with equation~\ref{lpvn} with (\cite{TimeE} 
equations~30~and~63): 
\pagebreak
\begin{equation}
 \label{lvfsq}
    L_4(\bv_{56})_{\mathrm{E}_7} \: = \:
   q(\mcX,\mcY,\alpha,\beta) 
    \: = \: 1
\end{equation}
 where $\bv_{56} \equiv (\mcX,\mcY,\alpha,\beta) \in F(\htho)$, 
 with $\mcX,\mcY \in \htho$  and $\alpha,\beta \inn \rrr$. 
Given the straightforward embedding of the subgroup $\esi$ with the  27-dimensional 
representation $\mathbf{27}$ and its complex conjugate
 $\mathbf{\overline{27}}$  corresponding respectively to the  $\mcX \inn \htho$ and 
$\mcY \inn \htho$ subcomponents of $F(\htho)$ in equation~\ref{lvfsq}, the main
 consequences of this $\ese$ augmentation follow directly from equation~\ref{esibits}. 
On now projecting the four components of
  $\bv_4 \equiv \bh\inn \htwc$ onto the local tangent space of the 
 external 4-dimensional spacetime from the $\mcY \inn \htho$ subcomponents the 
breaking pattern for the $\ese$ symmetry of equation~\ref{lvfsq} is determined 
(\cite{TimeE} subsection~4.3):
\begin{eqnarray}
        \ese \!\! &  \,\, \to \!\! &  \,\, 
		  \sltc \; \times \; \suth_c \; 
		   \times \; \uo_Q   \label{esesb} \\
	     \mcX\inn\htho \!\!\!\!\! & : &  \!\!\!\!\!\!   
    \left\{  \begin{array}{cccc}
       \, \mbox{vector}\,  \quad\;  & \quad\,  \mathbf{1} \quad\, & 
	                   \quad\;\;  0  \quad\; : &  \!\!\!\!   \mbox{`$\nu_L$'}  \\
	   \, \mbox{scalar}\,  \quad\;  & \quad\, \mathbf{3} \quad\, &
	          \quad\;\;  \frac{2}{3} \quad\;
			    \hspace{-0.5pt} : \hspace{0.5pt}  &  \!\!\!\! \mbox{`$u_L$'}  \\
       \! L\mbox{-spinor} \,  \quad\!\!\;  & \quad\,  \mathbf{1} \quad\, & 
	                       \quad\;\;  1  \quad\; : &  \!\!\!\!  e_L  \\
	   \! L\mbox{-spinor} \,  \quad\!\!\;  & \quad\, \mathbf{3} \quad\, &
	                       \quad\;\;  \frac{1}{3} \quad\;
			 \hspace{-0.5pt}   : \hspace{0.5pt}  &  \!\!\!\!  d_L   \\
		\, \mbox{scalar}\,  \quad\;  & \quad\, \mathbf{1} \quad\, &
	                       \quad\;\;  0  \quad\; : & \!\!\!\!	   n			   
	  \end{array}   \right.                  \nonumber                     \\
\raisebox{0pt}[0pt][0pt]{ {\raisebox{+9.3ex}{$\bv_{56} \to 
  \left\{ \begin{array}{c} \\ \\ \\ \\ \\ \\ \\ \\ \\ \\  \vspace{+1.0ex} \\
   \end{array}  \right. 
   \!\!\!\!\!\! $}} }		  
	 \mcY\inn\htho \!\!\!\!\! & : &  \!\!\!\!\!\!   
    \left\{ \hspace{-0.8pt} \begin{array}{cccc}
       \, \mbox{vector}\,  \quad\;  & \quad\,  \mathbf{1} \quad\, & 
	                       \quad\;\;  0 \quad\; : & \!\!\!\!   \bh  \\
	   \, \mbox{scalar}\,  \quad\;  & \quad\, \mathbf{3} \quad\, &
	                \quad\;\;  \frac{2}{3} \quad\;
		   \hspace{-0.7pt} : \hspace{0.7pt} & \!\!\!\! \mbox{`$u_R$'}  \\
       \! R\mbox{-spinor} \, \quad\!\!\;  & \quad\,  \mathbf{1} \quad\, & 
	                       \quad\;\;  1 \quad\; : & \!\!\!\!   e_R  \\
	   \! R\mbox{-spinor} \, \quad\!\!\;  & \quad\, \mathbf{3} \quad\, &
	                       \quad\;\;  \frac{1}{3} \quad\;
				 \hspace{-0.7pt} : \hspace{0.7pt} & \!\!\!\!  d_R  \\
		\, \mbox{scalar}\,  \quad\;  & \quad\, \mathbf{1} \quad\, &
	                       \quad\;\;  0 \quad\; : &	\!\!\!\!   N		   
	  \end{array}   \right.                  \nonumber                     \\ 
	     \alpha,\beta\inn \rrr   \!\!\!\!\! & : &    \quad\!
	   \hspace{-0.8pt}   \mbox{scalar}  \quad\, \qquad \;\;\:\!  \mathbf{1} \quad\, 
	                       \qquad\;\;   0 
		  \quad\; \hspace{-0.1pt} :   \:\!   \alpha,\beta  
						   \begin{array}{c} \vspace{0pt} \\ \end{array}  			   
						 \label{esebits}           
\end{eqnarray} 

   In addition to the four left-handed spinors of equation~\ref{esibits},
 reproduced in the $\mcX\inn\htho$ subcomponents above,
    a corresponding set of four \textit{right}-handed spinors is identified in the 
$\mcY\inn \htho$ subcomponents.
  Hence the $\mcX$ and $\mcY$ components of equation~\ref{esebits} are referred to as 
the `left-handed' and `right-handed' sectors of the theory. With the internal symmetry 
transformations being the same for both sectors, the \mbox{2-component} Weyl spinors 
for the $e$ and $d$ states in equation~\ref{esibits} have been augmented to 
4-component Dirac spinors in equation~\ref{esebits}. Corresponding $L$ and $R$ 
subscripts are also added to the `$u$-quark' and `$\nu$-lepton' states in 
equation~\ref{esebits}, albeit within quotation marks since the need to identify a 
Lorentz spinor structure for these states will require yet further augmentation.

 However we can observe at this stage that the embedding of the external 
\mbox{4-vector} $\bh\inn\htwc$, closely linked with the Higgs, within the $\mcY \in 
\htho$ components \textit{prohibits} the accommodation of a neutrino state `$\nu_R$' 
in the right-handed sector while implying that the slot is now \textit{open} for a 
left-handed neutrino `$\nu_L$' in the corresponding components of $\mcX \in \htho$, 
without the conflict described for equation~\ref{esibits}.  

  More generally 
   we note that the branching patterns obtained for this elementary symmetry breaking 
for natural augmentations of the form of time in equation~\ref{lpvn}
 over the local structure of \mbox{4-dimensional} spacetime,  leading to 
equations~\ref{slbits}, \ref{esibits} and \ref{esebits}, 
   provide a far better template
for the \textit{direct} emergence of the
Standard Model elementary particle multiplet structure 
than the equivalent analysis applied for the restricted case of extra spatial 
dimensions via the quadratic terms of equation~\ref{vnd}, as described for 
equations~\ref{sosb}--\ref{vnbits}, and with very little redundancy. 
   This strongly suggests that in place of equation~\ref{snd} the generalisation to 
equation~\ref{salpha} affords a more appropriate core basis for a unified theory, 
particularly since in the latter case we begin by discarding an
 assumption and
 the extraction of a necessarily quadratic substructure for external spacetime might 
underlie the \textit{mechanism} of symmetry breaking itself.

  Nevertheless further structure is still needed to describe the full Standard Model. 
  In addition to the required spinor structure for the $\nu$-lepton and $u$-quark  
states in equations~\ref{esesb}--\ref{esebits} the principle features that remain to 
be accounted for are an electroweak $\sutw_L \times \uo_Y$ symmetry (that breaks to 
$\uo_Q$) and a full three generations of leptons and quarks. Ideally a further 
\textit{natural} mathematical generalisation might incorporate these features. 
This leads to the proposal of a possible symmetry action of $\ee$, uniquely the 
largest exceptional Lie group,  on a homogeneous polynomial form:
\begin{equation}
 \label{lvto}
 \lvtfep
\end{equation}
 as the ultimate instantiation for equation~\ref{lpvn}.
  This provisional  form is potentially of octic order with $p=8$ (see for 
example~\cite{CedP}), and a close connection with the smallest non-trivial $\ee$ 
representation with $n=248$ dimensions is here presumed; although other values for $p$ 
and $n$ might be conceivable.
  The nature of this structure and  
 the plausibility of encompassing the principle remaining Standard Model features  in 
a correlated manner is the main topic of~\cite{TimeE}.

As a unification symmetry the Lie group $\ee$ itself is comfortably able to 
incorporate a broken symmetry  
 corresponding to a product of the external Lorentz group and internal Standard Model 
gauge groups in the form of equation~\ref{gbreak} with:
\begin{equation}
 \label{eebrk}
    \mbox{Lorentz} \, \times \, \suth_c \times \sutw_L \times 
	           \uo_Y \, \subset \, \ee
\end{equation}	
 On the other hand as an extension from the $\mathbf{27}$ representation of $\esi$ 
underlying equation~\ref{esibits}, as
 combined with the complex conjugate $\mathbf{\overline{27}}$ for  
equation~\ref{esebits}, 
	     a possible factor of three for three generations of leptons and quarks is 
suggested by the subgroup embedding of $\esi$ in $\ee$ with the representation 
branching pattern:
\begin{equation}
  \label{eetoesi}
  \ee \supset \esi \times \suth : \quad
    \mathbf{248} \, \to \, 
	(\mathbf{27},\mathbf{3}) + (\mathbf{\overline{27}},\mathbf{\overline{3}})
	+ (\mathbf{78},\mathbf{1}) + (\mathbf{1},\mathbf{8}) 
\end{equation}	 

  However, as explained in~\cite{TimeE}, unlike the case for the embedding of the 
$\esi \subset \ese$ action the embedding of  $\esi$ and $\ese$  in the $\ee$ action on 
$\lvtfep$ is expected to be less direct than that suggested by equation~\ref{eetoesi} 
if the needed spinor structures for $\nu$-leptons and $u$-quarks and a complete 
electroweak theory are also to be identified compatible with the symmetry breaking 
pattern of equation~\ref{eebrk}.
 As a generalisation from the $\esi$ action on equation~\ref{lvts} and the $\ese$ 
action on equation~\ref{lvfsq} the proposed $\ee$ action is also presumed to 
incorporate octonion composition in an essential way, with the properties of octonion 
`triality' (\cite{TimeE} section~5) expected to be at the heart of unravelling the 
full Standard Model spinor structure for a full three generations of leptons and 
quarks. 

\section{Neutrinos and other New Physics}
\label{una6}

   Notwithstanding the above caveat regarding the need to incorporate a neutrino 
spinor structure, given the embedding of the $\esi$ level of equation~\ref{esibits} 
within the $\ese$ level of equation~\ref{esebits} we might anticipate some of the 
implications for neutrino physics  
of a further embedding within a three generation pattern at the $\ee$ level, if we 
assume in broad terms the simplest further progression for the neutrino sector:
\begin{eqnarray}
 \esi \!\!\! & : &  \begin{tabular}{|c|} \hline 
        \raisebox{+0.25ex}{$\nu_L / \bh$} \\ 
                                                        \hline \end{tabular}
\qquad\qquad\qquad\qquad\qquad\qquad\;\;\;
               \mbox{(equation~\ref{esibits})}	 \nonumber  \\
 \ese \!\!\! & : &  \begin{tabular}{|c|} \hline 
        \raisebox{+0.25ex}{$\nu_L$}          \\ \hline \end{tabular} 
               \qquad \qquad \;\,\, \mbox{and} \quad
	           \begin{tabular}{|c|} \hline 
	    \raisebox{+0.25ex}{$\:\bh\:$} 	    \\ \hline \end{tabular}                    
   \qquad\qquad\;\;    \mbox{(equation~\ref{esebits})}	   \nonumber  \\  
 \ee \!\!\! & : &  \begin{tabular}{|c|} \hline 
        \raisebox{+0.25ex}{$\nu_L \;\;\, \nu_L \;\;\, \nu_L$}  
		                    \\ \hline \end{tabular}
	           \quad \mbox{and} \quad
	           \begin{tabular}{|c|} \hline 
		\raisebox{+0.25ex}{$\:\bh\:\;\;\, \nu_R \;\;\,\nu_R$} 	   
			    \\ \hline \end{tabular} 
			    \label{eebits}		
\end{eqnarray}

  This schematic augmentation incorporates  three generations of left-handed 
neutrinos, as for the left and right-handed charged leptons and quarks, and  suggests 
the accommodation of \textit{two right-handed} neutrinos alongside the original 
external $\bh \inn \htwc$ projection, which now prohibits the identification of a 
third $\nu_R$ state.
 With the components of $\bh \equiv \bv_4 \in \TM_4$ being associated with the Higgs 
and the origin of mass (as discussed before and following equation~\ref{gbreak})
 a clear origin for a mass asymmetry in the neutrino sector is also implied 
 for this provisional structure at the $\ee$ level in equation~\ref{eebits}, which 
suggests some form of `seesaw' imbalance between the left and right-handed states.  
 In a standard neutrino seesaw 
 mechanism model (see for example~\cite{Drewes} section~2 and references therein)
 each $\nu_R$ state generates one $\nu_L$ mass. Hence
 with only two $\nu_R$ states available in the provisional scheme of 
equation~\ref{eebits} there is an indication that the lightest $\nu_L$ mass state may 
in fact be massless, that is $\mmin = 0$.

 While ongoing and future experiments on tritium beta decay~\cite{Katrin} and 
neutrinoless double-beta decay~\cite{Agost} will improve the corresponding constraints 
on the mass of the lightest $\nu_L$ state the most stringent test of a predicted 
$\mmin = 0$ may be provided by the cosmological observations limiting  the total mass 
of the three $\nu_L$ states,  currently with an upper bound of around 
$\mtot < 0.20\,\mbox{eV}$~(\cite{PDG18} section~64). 
  Given the two empirically established $\nu_L$ mass differences from  
 solar and atmospheric neutrino oscillations the lowest possible value is   $\mtot 
\simeq 0.06\,\mbox{eV}$  (see also~\cite{Drewes} section~2). Although model-dependent, 
future prospects for standard neutrinos within the $\Lambda$CDM cosmological model
  (with $\Lambda$ the cosmological constant and CDM cold dark matter)
  are for $\mtot = 0.06\,\mbox{eV}$ to be detectable at the 3--4$\sigma$ level in the 
coming years~(\cite{PDG18} sections~25.4 and 64).
 This implies that the case for $\mmin = 0$ is testable in that it could be 
disfavoured with statistical significance in the foreseeable future.

 Without the guide of equation~\ref{eebits} 
  a more symmetric proposal would be for the introduction of \textit{three} $\nu_R$ 
states  as for the `Neutrino Minimal Standard Model',
  or $\nu$MSM~(\cite{Drewes} section~7, \cite{Asaka1,Asaka2}), proposed as a simple 
economical extension from the Standard Model (for which all three $\nu_L$ states are 
massless and there are no $\nu_R$ states).
 Two of the $\nu_R$ states in the $\nu$MSM  have nearly degenerate masses in the range 
of around 1--100$\,\mbox{GeV}$ which generate $\nu_L$ masses via the seesaw mechanism 
consistent with the solar and atmospheric neutrino oscillation data and also in 
principle accounting for the baryon asymmetry of the universe through $C\!P$-violating 
oscillations of these two $\nu_R$ states in its early history. 
   Hence the $\nu$MSM implies that the two $\nu_R$ states in equation~\ref{eebits} may 
also be sufficient to account for these phenomena. 

The third $\nu_R$ of the $\nu$MSM, with a mass of a few keV, acts as a warm dark 
matter candidate but with a Yukawa coupling too small to make a significant 
contribution to the $\nu_L$ masses and hence leaving the lightest $\nu_L$ state 
practically massless.
   However here from equation~\ref{eebits} there is no room to accommodate a third 
$\nu_R$ state, suggesting that $\mmin=0$, and a third $\nu_R$ is not needed for dark 
matter since such candidates are provided by the scalar invariant components 
$n,N,\alpha$ and $\beta$ listed in equation~\ref{esebits}.
 These augment the original single scalar invariant $n\in \rrr$ of 
equations~\ref{slbits} and \ref{esibits}.  The four scalar invariants at the $\ese$ 
level in equation~\ref{esebits} may generalise into a broader `dark sector' involving 
  further components at the full $\ee$ symmetry level and offer the possibility to 
observationally test this new physics by exploring the corresponding cosmological 
implications.

 From equation~\ref{eebits} the Standard Model Higgs, deriving from the components 
underlying  $\bh \equiv \bv_4 \in \TM_4$, is clearly intimately connected with the 
neutrino sector, and hence some of their properties may be closely correlated. We note 
for example that at the upper end of the
  1$\sim$100$\,\mbox{GeV}$  range suggested by the $\nu$MSM the mass scale for two
   $\nu_R$ states   would be close to the observed value of $M_H \simeq 
125\,\mbox{GeV}$ for the Higgs~\cite{PDG18}. 
  The need for a spinor structure for both the $\nu_L$ and $\nu_R$ states to be 
incorporated under the $\mbox{Lorentz} \subset \ee$ action of equation~\ref{eebrk} on 
the components of $\lvtfep$ for
equation~\ref{eebits} suggests that $\bh$ may itself also have an underlying spinor 
composition. Indeed spinor components can be combined to form both vector objects, as 
for $\psi\psi^{\dagger}$ in equation~\ref{lqinc},
  and scalar objects as for composite models for the Higgs (see for 
example~\cite{Azat} and references therein).
 For composite Higgs models the coupling of the Higgs to fermion pairs can deviate 
from the Standard Model expectation by of order 10\%, sufficient for this new physics 
to be observable at a 250$\,$GeV $e^+e^-$ linear collider 
 (\cite{Peskilc} section 5). At such a machine invisible Higgs decays to a dark or 
hidden sector, such as the $n,N,\alpha,\beta$ or further scalar invariant components 
discussed above, can also be inferred via a visible recoiling $Z^0$ boson decay
  (\cite{Peskilc} section 6).

  Precise empirical predictions 
 will require a full understanding of the structure of the theoretically predicted 
$\ee$ symmetry action on the full form of time $\lvtfep$  and the resulting symmetry 
breaking pattern.
   Since the further required features of the Standard Model beyond
    equations~\ref{esesb}--\ref{esebits} are closely correlated it is plausible that 
they may all be uncovered together in one further augmentation from the $\ese$ form of 
equation~\ref{lvfsq} to the proposed $\ee$ form described for equation~\ref{lvto} (as 
considered in detail in~\cite{TimeE}). If these required features emerge at the $\ee$ 
level this will provide a very firm basis for investigating a wealth of new physics 
beyond the Standard Model.


\section{Interpretation and Opportunities}
\label{una7}

While connecting with current and future programs in both experimental particle 
physics and observational cosmology the theory presented here can be motivated in a 
similar spirit as for the early unified field theories dating from a hundred years ago  
briefly discussed in the opening of this paper. 
Following those initial proposals, in the early 1930s and still early in his personal 
quest for such a theory, Einstein summed up the nature of the problem of seeking a 
unifying extension to general relativity with the question~\cite{Eincont}:
\pagebreak
 \begin{quotation}
    Is there a theory of the continuum in which a new structural element 
	appears side by side with the metric such that it forms a single whole
   together with the metric?
 \end{quotation}
 
    Here equation~\ref{salpha}, equivalent to equation~\ref{lpvn}, is proposed as such 
a `single whole', deriving from the continuum of proper time, which can incorporate 
the local 4-dimensional spacetime metric, as described for equation~\ref{sfourp} and 
exemplified by the cubic form of equations~\ref{lvni}--\ref{lqinc}, side by side with 
additional structures that are interpreted as a basis for matter fields in spacetime.

Historically in progressing from a Newtonian absolute space and absolute time to the 
spacetime of special and then general relativity in the early 
 $20^{\mathrm{th}}$ century a central role was played by the conception of time,
 with a different proper time carried at rest in each inertial reference frame.
  In general relativity the inertial frames are strictly local with each proper time 
interval $\delta s$ invariant under local Lorentz transformations and taking the form 
of equation~\ref{sfourd}. On the global scale the Lorentz metric $\eta_{ab}$ is  
supplanted by the general metric $g_{\mu\nu}(x)$, identified with the gravitational 
field, with the local proper time interval in equation~\ref{gfourd} invariant under 
general coordinate transformations.

 It is this distinctive invariant role for time that we have focussed upon and 
generalised in leading from equation~\ref{sfourd}, via equation~\ref{snd}, to 
equations~\ref{salpha} and \ref{lpvn}. 
  Matter fields originate at the most elementary level through a simple symmetry 
breaking analysis for equation~\ref{lpvn},
  deriving directly from the extraction of a $\mbox{Lorentz} \subset \hG$ external 
symmetry acting on the  subcomponents $\bv_4 \in \TM_4$ projected
 from $\bv_n \in \rrr^n$ onto the local external \mbox{4-dimensional} spacetime. 
 Through natural augmentations the properties of the residual components are found to 
resemble structures of the Standard Model
  at the elementary particle level of matter as described for equations~~\ref{slbits}, 
\ref{esibits} and \ref{esebits}, 
 without needing to contrive or postulate these features.
 The contrast with general relativity, which considers the \textit{global} geometry of 
spacetime through equation~\ref{gfourd} rather than the purely \textit{local} 
generalisation for proper time $\delta s$ from equation~\ref{sfourd} to 
equation~\ref{salpha}, is depicted in figure~\ref{grcfme}.
\vspace{-2pt}
\begin{figure}[htbp]  
\centering
\epsfxsize=14.25cm
\leavevmode
\epsffile[0 0 1669 729]{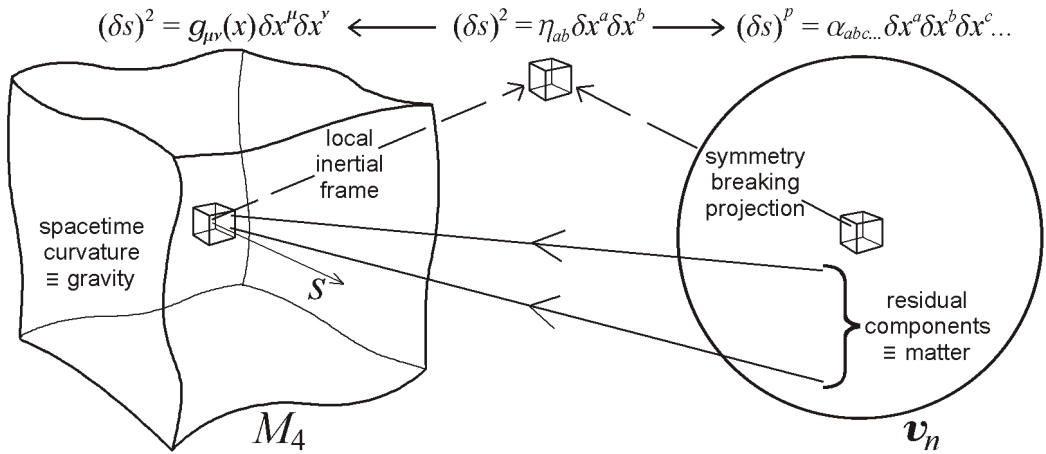}
\vspace{-10pt}
\caption{\setb
The relation between gravitation, described  by the metric field $g_{\mu\nu}(x)$ on 
$M_4$ in generalising from the \textit{global} Lorentz metric $\eta_{ab}$ of a flat 
4-dimensional spacetime, 
and matter fields, originating from the broken symmetry of
  a complementary direct generalisation of the 
 \textit{local} Lorentz metric expression for a
  proper time interval $\delta s$.}
\label{grcfme}
\end{figure}  

As implied in figure~\ref{grcfme} we can zoom into an infinitesimal local inertial 
reference frame anywhere on the spacetime manifold $M_4$ and generalise 
equation~\ref{sfourd} to equations~\ref{salpha} and \ref{lpvn} to explore the 
microscopic structure of matter deriving from the residual components.
 The theory is essentially founded upon
  the continuous flow of time that can be associated with any local inertial frame at 
any location in spacetime.
  Analysing this simple `one-dimensional' starting point a finite duration of proper 
time $s \in \rrr$  can be decomposed down to infinitesimal intervals:
\begin{equation}
  \label{sdecom}
  s = \delta s + \delta s +  \delta s +  \ldots
    \quad \mbox{with substructure} \quad\;\;
  \delta s =  {}^{\substack{ p \vspace{1pt} \\ {} }} \!\!\!\! 
    \sqrt{\alpha_{abc\ldots}\delta x^a \delta x^b \delta x^c \ldots}	 
\end{equation} 
 for $\delta s \to 0$ as the $p^{\mathrm{th}}$-root of a homogeneous polynomial of 
order $p$ in $n$ components $\{\delta x^a\} \in \rrr^n$; with $a,b,c = 1,\ldots ,n$
 (or = $0,\ldots,n-1$) and each  $\alpha_{abc\ldots} \inn \{-1,0,1\}$. Here we are 
simply exploiting the basic arithmetic properties of the real numbers, representing 
the continuum of time with $\delta s \in \rrr$, which together with addition include 
the operations of multiplication and extracting roots. The right-hand expression in 
equation~\ref{sdecom} is precisely equivalent to equation~\ref{salpha}, which as a 
natural generalisation from the expression for extra spatial dimensions in 
equation~\ref{snd} forms the basis of the whole theory.  
  In equations~\ref{sdecom} and \ref{salpha} we are not \textit{adding} anything to 
time, nor \textit{replacing} time with anything, but simply expressing an intrinsic 
arithmetic substructure that is carried \textit{simultaneously} with time. 
 The simplicity of this interpretation then provides a further motivation for the 
theory 
 (the historical and philosophical aspects of which are discussed in~\cite{Struct}).
 
  In summary there are three main supporting arguments for this theory:
\begin{itemize}
  \item The \textit{assumption} that further components augmenting the local 
4-dimensional spacetime form for a proper time interval should have a spatial 
structure can be \textit{dropped}, provided that we can identify a 4-dimensional 
quadratic spacetime substructure from this generalisation as described for 
equations~\ref{sfourp} and \ref{lqinc}. The generalisation to cubic and higher order 
expressions is permitted since \textit{we do not perceive} the additional components 
as extra spatial dimensions.

  \item An underlying \textit{simplicity} is achieved with the theory interpreted as 
deriving from the continuum of time alone, as described for equation~\ref{sdecom}. 
With potentially no further substructure either possible or needed, and with the flow 
of time infusing all experiments and observations, there is a suggestion of reaching 
the ultimate `bedrock' underlying the structure of matter as observed in spacetime, 
rather than the next of an indefinite sequence of substrata.

 \item Significant connections with the \textit{empirical} properties of the Standard 
Model are obtained from the natural mathematical development of the theory, including 
the  identification of Lorentz spinors, colour $\suth_c$ singlets and triplets with 
the appropriate electromagnetic $\uo_Q$ fractional charges and an intrinsic left-right 
asymmetry, which is particularly marked for the neutrino sector, and with very little 
redundancy of structure as described for equation~\ref{esebits}.
\end{itemize}

  These elements of the Standard Model have been accounted for through a rigorous 
analysis of the $\esi$ and $\ese$ levels underlying 
equations~\ref{lvts}--\ref{esebits}.  The technical mathematical details are described 
extensively in~(\cite{TimeE} and references therein) and 
 naturally lead to the prediction of an $\ee$ symmetry of the form of time described 
for equation~\ref{lvto} to uncover the 
full particle multiplet structure of the Standard Model and beyond.
  In particular the mathematical pursuit of a full symmetry action for $\ee$ on the 
form $\lvtfep$ and the resulting breaking pattern may elucidate the origin of the 
esoteric properties of neutrinos, building upon the schematic generation structure of 
equation~\ref{eebits}. Mutually, existing empirical knowledge 
of the neutrino sector as established in recent decades, as expressed for example in 
the $\nu$MSM, might also be utilised
 as a key component in unlocking the detailed structure of the specific application 
for $\ee$ that is proposed. 

   More generally the mathematical possibilities for equation~\ref{lpvn} 
are open to exploration, as are the wider implications of the theory in relation to 
Kaluza-Klein models and quantisation schemes on the technical side and cosmology and 
particle physics on the empirical side.   
  The manner in which this theory has been able to reproduce a series of basic 
features of the Standard Model, and has already yielded provisional connections with 
neutrino and other new physics beyond, demonstrates this open opportunity to further   
develop this fundamental unified theory
    in parallel with advances in our empirical
	understanding of the elementary composition of the physical world.




{\setlength{\baselineskip}{0.9\baselineskip}

\par}



\par}


\begin{thebibliography}{99}

\bibitem{Pais2}
  Abraham Pais,
  `Inward Bound: Of Matter and Forces in the Physical World', 
  Oxford University Press (1986, 1988, 1994).
  
\bibitem{Pais} 
  Abraham Pais, 
  ` ``Subtle is the Lord\ldots'': The Science and the Life of Albert Einstein', 
  Oxford University Press (1982,~2005). 

\bibitem{Eingr}  
  Albert Einstein,
  `The Foundation of the General Theory of Relativity',
  Annalen Phys.\  {\bf 49} (7), 769--822 (1916)
  [Annalen Phys.\  {\bf 14} (Supplement), 517--571 (2005)],
  translated by Alfred Engel in `The Collected Papers of Albert Einstein' {\bf 6},
  146--200, Princeton University Press (1997).   

\bibitem{Weyl1}
  Hermann Weyl,
  `Gravitation and Electricity',
  Sitzungsber.\ Preuss.\ Akad.\ Wiss.\ Berlin (Math.\ Phys.) {\bf 1918}, 465 (1918).
  
\bibitem{ORaif}
  L.~O'Raifeartaigh,
  `The Dawning of Gauge Theory', 
   Princeton University Press (1997).

\bibitem{Weyl3}
  Hermann Weyl,
  `Electron and Gravitation',
   Z.\ Phys.\  {\bf 56}, 330 (1929)
  [Surveys High Energ.\ Phys.\  {\bf 5}, 261 (1986)].
	 
\bibitem{Gur1} 
  F.~G\"{u}rsey, P.~Ramond and P.~Sikivie, 
  `A Universal Gauge Theory Model based on $\esi$',
  Phys.\ Lett.\ B {\bf 60} (2), 177--180 (1976).

\bibitem{Gur7} 
  F.~G\"{u}rsey and P.~Sikivie, 
  `$\ese$ as a Universal Gauge Group',
  Phys.\ Rev.\ Lett.\ {\bf 36} (14), 775--778 (1976).
  
\bibitem{Kaluza}
  T.~Kaluza,
  `On the Problem of Unity in Physics',
  Sitzungsber.\ Preuss.\ Akad.\ Wiss.\ Berlin (Math.\ Phys.) {\bf 1921}, 966 (1921).
   
\bibitem{Klein}
  O.~Klein,
  `Quantum Theory and Five-Dimensional Relativity',
    Z. Phys. \textbf{37}, 895 (1926).
  
\bibitem{Cho} 
  Y.~M.~Cho,
  `Higher-Dimensional Unifications of Gravitation and Gauge Theories',
  J.\ Math.\ Phys.\  {\bf 16} (10), 2029 (1975).
  
\bibitem{KKone}
  D.~J.~Jackson,
  `Construction of a Kaluza-Klein type Theory from One Dimension',
   arXiv:1610.04456 [physics.gen-ph] (2016).   
  
\bibitem{TimeE}  
  D.~J.~Jackson,
  `Time, $\ee$, and the Standard Model',
  arXiv:1709.03877 [physics.gen-ph] (2017).
  
\bibitem{Baez1} 
  J.~C.~Baez,
  `The Octonions',
  Bull.\ Am.\ Math.\ Soc.\  {\bf 39}, 145--205 (2002)
  [arXiv:math/0105155 [math.RA]].
  
\bibitem{Chev}
  Claude Chevalley and Richard D. Schafer, 
  `The Exceptional Simple Lie algebras $\ff$ and $\esi$', 
   Proc.\ Natl.\ Acad.\ Sci.\ USA\ \textbf{36}, 137--141 (1950).   
   
\bibitem{Wang2} 
   A.~Wangberg and T.~Dray, 
  `$\esi$, the Group: The Structure of $\sltho$', 
  J.\ Algebra\ Appl.\ {\bf 14} (6), 1550091 (2015)
  [arXiv:1212.3182 [math.RA]].    
  
\bibitem{Freud} Hans Freudenthal, 
  `Lie Groups in the Foundations of Geometry', 
   Adv.\ Math.\ \textbf{1}, 145--190 (1964).
   
\bibitem{Rios} 
  M.~Rios,
  `Jordan $C^{\ast}$-Algebras and Supergravity',
  arXiv:1005.3514 [hep-th] (2010).
  
\bibitem{CedP} 
  M.~Cederwall and J.~Palmkvist, 
  `The Octic $\ee$ Invariant',
  J.\ Math.\ Phys.\  {\bf 48}, 073505 (2007)
  [arXiv:hep-th/0702024].
  
\bibitem{Drewes}
   M.~Drewes,
  `The Phenomenology of Right Handed Neutrinos',
  Int.\ J.\ Mod.\ Phys.\ E {\bf 22}, 1330019 (2013)
  [arXiv:1303.6912 [hep-ph]].
  
\bibitem{Katrin}
   J.~Wolf [KATRIN Collaboration],
   `The KATRIN Neutrino Mass Experiment',
   Nucl.\ Instrum.\ Meth.\ A {\bf 623}, 442--444 (2010)
  [arXiv:0810.3281 [physics.ins-det]].

\bibitem{Agost}
   M.~Agostini, G.~Benato and J.~Detwiler,
   `Discovery Probability of Next-Generation Neutrinoless Double-$\beta$ Decay Experiments',
   Phys.\ Rev.\ D {\bf 96} (5), 053001 (2017)
  [arXiv:1705.02996 [hep-ex]].
  
\bibitem{PDG18}
   M.~Tanabashi {\it et al.} [Particle Data Group],
   `Review of Particle Physics',
   Phys.\ Rev.\ D {\bf 98} (3), 030001 (2018),
   [available at http://pdg.lbl.gov].
	
		
\bibitem{Asaka1} 
  T.~Asaka, S.~Blanchet and M.~Shaposhnikov,
  `The $\nu$MSM, Dark Matter and Neutrino Masses',
  Phys.\ Lett.\ B {\bf 631}, 151 (2005)
  [arXiv:hep-ph/0503065].
  
\bibitem{Asaka2} 
  T.~Asaka and M.~Shaposhnikov,
  `The $\nu$MSM, Dark Matter and Baryon Asymmetry of the Universe',
  Phys.\ Lett.\ B {\bf 620}, 17 (2005)
  [arXiv:hep-ph/0505013].
    
\bibitem{Azat}
   A.~Azatov,
   `Status of Composite Higgs',
  PoS EPS{\bf -HEP2017}, Venice (5--12 July 2017), 255 (2017)
  [doi:10.22323/1.314.0255].
 
\bibitem{Peskilc}
  K.~Fujii {\it et al.},
  `Physics Case for the 250$\,$GeV Stage of the International Linear Collider',  
  arXiv:1710.07621 [hep-ex] (2017).
  
\bibitem{Eincont}
   Albert Einstein,
   `The Problem of Space, Ether, and the Field in Physics',
   essay in the volume `Mein Weltbild', 
   translated by Alan Harris in `The World as I see it',
   John Lane The Bodley Head, London (first published in England in 1935). 
   [The passage quoted appears in the first of the final seven paragraphs which
    were typically cut for later volumes incorporating this essay.]

\bibitem{Struct}  
  D.~J.~Jackson,
  `The Structure of Matter in Spacetime from the Substructure of Time',
  arXiv:1804.00487 [physics.gen-ph] (2018).   
    
  
\end{thebibliography}
\end{document}